\documentstyle[aps,prc,twocolumn,floats,epsfig]{revtex}
\draft
\renewcommand{\vec}[1]{{\mathbf{#1}}}
\begin{document}

\renewcommand{\thefootnote}{\arabic{footnote}}

\twocolumn[\columnwidth\textwidth\csname@twocolumnfalse\endcsname

\title{Beta decay of r-process waiting-point nuclei in a self-consistent approach}

\author{J. Engel,$^{1}$
        M. Bender,$^{1,2}$
        J. Dobaczewski,$^{2,3,4}$
        W. Nazarewicz,$^{2,3,5}$ and
        R. Surman$^{1}$
}

\address{${}^{1}$Department of Physics and Astronomy,
         The University of North Carolina,
         Chapel Hill, North Carolina 27599-3255}

\address{$^2$Department of Physics and Astronomy,
         University of Tennessee,
         Knoxville, Tennessee 37996}

\address{$^3$Institute of Theoretical Physics,
         Warsaw University,
         Ho\.za 69,  PL-00681, Warsaw, Poland}

\address{$^4$Joint Institute for Heavy Ion Research,
         Oak Ridge National Laboratory,
         P.O. Box 2008, Oak Ridge, Tennessee 37831}

\address{$^5$Physics Division, Oak Ridge National Laboratory,
         Oak Ridge, Tennessee 37831}

\date{February 19 1999}

\maketitle

\addvspace{5mm}

%
%
\begin{abstract}
Beta-decay rates for spherical neutron-rich r-process waiting-point
nuclei are calculated within a fully self-consistent Quasiparticle
Random-Phase Approximation, formulated in the The Hartree-Fock-Bogolyubov
canonical single-particle basis. The same Skyrme force is used
everywhere in the calculation except in the proton-neutron particle-particle
channel, where a finite-range force is consistently employed.  In all but the
heaviest nuclei, the resulting half-lives are usually shorter by factors of 2
to 5 than those of calculations that ignore the proton-neutron
particle-particle interaction. The shorter half-lives alter predictions for
the abundance distribution of r-process elements and for the time it takes to
synthesize them.

\end{abstract}
\pacs{PACS numbers:
      21.30.Fe, 
      21.60.Jz, 
      26.30.+k, 
      23.40.-s, 
      23.40.Hc  
}

\addvspace{5mm}]

\narrowtext
%
%

\section{Introduction}
\label{Sect:intro}

The astrophysical r-process \cite{burbidge,clayton,cowan,MeM92,Kratz,WiT93},
which creates about half of all nuclei with \mbox{$A > 70$}, proceeds through
very neutron-rich and unstable isotopes produced by stellar explosions or
other violent events.  The
ultimate abundance of any stable nuclide depends strongly on the beta-decay
half lives of its neutron-rich progenitors.  The solar elemental abundance
distribution shows peaks near $A$=80, 130, and 195, corresponding to
progenitors with closed neutron shells ($N$=50, 82, and 126).  These
relatively long-lived nuclei not only define the abundance peaks but also
restrict the amount of heavier material that is synthesized.  Understanding
the important features of the r-process therefore requires knowledge of
lifetimes of closed-shell semimagic nuclei far from stability.  Of course beta
decay is only one of the processes that contribute to r-process abundances;
neutron capture and photodisintegration also play important roles, as do the
temperature, density and initial neutron to seed ratio in the explosive
environment.  But these other aspects are beyond the scope of this
article, which focuses on the crucial question of how to calculate beta decay
far from stability.  At present, the very neutron-rich nuclei are out of
experimental reach, and theory provides the only handle on their decay rates.

Precise theoretical estimates of beta-decay rates are hard to make.  Most of
the strength associated with the $\beta^-$-decay operator $\vec{\sigma}
\tau_+$ lies in the Gamow-Teller (GT) resonance, well above decay threshold.
The strength that actually contributes to beta decay is the small low-energy
tail of the GT distribution.  Calculated beta-decay rates can therefore vary
over a wide range without coming into conflict with sum rules, which for other
processes help reduce theoretical uncertainty.  In addition, beta-decay
lifetimes depend sensitively on nuclear binding energies, and small errors in
calculated $Q_\beta$-values can cause large errors in predicted decay rates.
These problems are complicated enough to demand a coherent and systematic
approach to beta decay.  Here we make a first attempt at a completely
self-consistent calculation.  Our goal, in spite of the difficulties, is a
reliable estimate of beta-decay rates in important even-even semimagic nuclei
lying on the r-process path.

Special tools are needed to describe transitions to low-lying excited states
in weakly bound nuclei.  Although large-scale shell model calculations
successfully describe the GT strength distribution in medium-mass nuclei close
to the valley of beta stability \cite{radha,martinez}, large configuration
spaces and difficulties with the continuum \cite{DN98} make the approach hard
to apply along the r-process path.  The continuum random phase approximation
\cite{[Mig67],[Shl75]} is often useful, but in very-weakly bound nuclei
pairing is important and a quasiparticle random phase approximation (QRPA)
based on coordinate-space Hartree-Fock-Bogolyubov (HFB) theory is required.
Surprisingly little work has been done along these lines.  Much more common
are global (in that they attempt to describe all nuclei in the same framework)
but non self-consistent calculations \cite{mol,Homma,hir,hir1} that substitute
the Strutinsky method + BCS for HFB, approximate the continuum by bound or
quasibound orbits, and use a schematic interaction
$\kappa(\vec{\sigma}_1\vec{\sigma}_2)(\vec{\tau}_1\vec{\tau}_2)$ in the QRPA.
The work of Ref.~\cite{zyl}, which successfully reproduced the half-lives of
the nickel isotopes and of $^{132}$Sn in the Tamm-Dancoff approximation, used
self-consistent single-particle energies and orbits but retained the
traditional schematic residual interaction.  But like the global calculations,
this work did not include a proton-neutron ($pn$) particle-particle
interaction in the QRPA.

Borzov {\it et al.}  \cite{Bor95,Bor96} did use a more self-consistent method
in restricted regions of the nuclide chart.  Their starting point was an
energy-density functional optimized for the regions they considered, and a
realistic interaction (including a zero-range $pn$ particle-particle
component) in the QRPA.  Their energy functional was spin-independent,
however, and so they did not obtain the residual interaction from the second
derivative of the functional, as required if the QRPA is to represent the
small amplitude limit of time-dependent HFB (this is what we mean by
self-consistency in an HFB + QRPA calculation).  Instead they chose a
phenomenological spin-dependent residual interaction \cite{Bor95}.  In
addition, they neglected positive-energy single-particle states in the HFB
calculation, though they did include the entire particle-hole continuum in the
QRPA.  In any event, Borzov {\it et al.}  \cite{Bor97,BorzovTours} have now
abandoned attempts at fully self-consistent calculations in favor of the
Extended Thomas-Fermi with Strutinsky Integral (ETFSI) approach.  No previous
papers, in sum, have used the same interaction in the mean-field calculation
and the QRPA, and most have neglected the residual $pn$ particle-particle
interaction.  Neither has any previous work properly included the effects of
the low-energy continuum on pairing.

Here we attempt to do better:  we use the same interaction in the HFB and QRPA
calculations so as to preserve the small-amplitude limit of time-dependent
HFB.  In the $pn$ particle-particle channel of the QRPA, we use a finite-range
interaction to soften the divergences induced by a delta function, but will
show that this choice does not spoil self-consistency.  By adjusting one
parameter, the strength of the $pn$ particle-particle
interaction (which we will
often refer to as $T$=0 pairing), we reproduce measured decay rates near the
closed neutron shells, and then make predictions for rates further from
stability.  For now we restrict our study to semi-magic nuclei with closed
neutron shells and their neighbors; besides being the most important for the
r-process these nuclei are conveniently spherical.

Our paper is organized as follows:  In Sec.~\ref{Sect:frame} we review the the
general formalism for calculating beta decay, briefly present coordinate-space
HFB theory and the QRPA, and then show how to combine them in a
self-consistent way.  Section \ref{Sect:inter} discusses the problem of
choosing a good effective interaction.  Section \ref{Sect:results} contains
the results of numerical calculations and an r-process simulation, and
Sec.~\ref{Sect:summary} presents conclusions.

%
%
\section{Framework}
\label{Sect:frame}
%
%
\subsection{Calculation of half-lives}
\label{Sect:halflive}

The rate for allowed Gamow-Teller decay of an even-even nucleus is given by
\begin{equation}
\label{eq:Fermi}
\frac{1}{T_{1/2}}
= \frac{2 \pi}{\ln 2}\frac{g_A^2}{\hbar} \int \! dE_e
  \sum_{m} | \langle 1^+_m | \sum_n \vec{\sigma}_n \tau^+_n | 0^+ \rangle |^2
  \; \frac{d n_m}{dE_e}
\ \ ,
\end{equation}
where the index $n$ labels particles, $|0^+\rangle$ is the initial ground 
state and the
$| 1^+_m \rangle$ are the final states (We use units in which $c$=1.)
The sum over $m$ runs over all final 1$^+$ states with an excitation 
energy smaller
than the $Q_{\beta^-}$ value.  We will set the weak axial nucleon coupling
constant $g_A$ to 1 rather than its actual value of 1.26 to account for the
near universal quenching of isovector spin matrix elements \cite{boh} in
nuclei. The kinematic factor $dn_m/dE$ is the density of final ($e^-,
\bar{\nu}_{e}$) states. It can be written as
\begin{equation}
\label{eq:spectrum}
\frac{dn_m}{dE_e}
= \frac{ E_{e} \; \sqrt{E_e^2 - m_e^2} }{2 \pi^3}
  \; ( E_i - E_{1^+_m} - E_{e} )^2 \; F (Z,E_{e})
\ \ ,
\end{equation}
where $E_i$ is the ground state energy of 
the even-even initial
nucleus with $Z$ protons and $N$ neutrons, $E_{1^+_m}$ is the energy of the
$m$th excited 1$^+$ state in the final nucleus with $Z+1$ protons and 
$N-1$
neutrons, and $F(Z,E_{e})$ corrects the
phase-space factor for the nuclear charge and finite nuclear size, both of
which affect the electron wave function.

We would like an approximation for $E_i - E_{1^+_m}$ in Eq.\
(\ref{eq:spectrum}) that does not require an explicit calculation of the
{$Q_{\beta^-}$} value.  To obtain one we express the
$Q_{\beta^-}$ value in terms of
nuclear ground-state binding energies:
\begin{equation}
\label{eq:binding}
Q_{\beta^-}
= \Delta M_{n-H} + B_{\rm g.s.}(Z,N) - B_{\rm g.s.}(Z+1,N-1)
\end{equation}
where $\Delta M_{n-H}$=0.78227\,MeV is
the mass difference between the neutron and the hydrogen atom.  The 
ground-state binding
energy of  an odd-odd final nucleus, in the independent quasiparticle
approximation, is given by
\begin{equation}
\label{oneplus}
B_{\rm g.s.}(Z+1,N-1)
\approx B_{\rm g.s.}(Z,N) + \lambda_p - \lambda_n + E_{2qp,{\rm lowest}}
\end{equation}
where $\lambda_p$$\equiv$$dE/dZ$ and $\lambda_n$$\equiv$$dE/dN$
are the proton and neutron Fermi energies in the initial nucleus ($Z,N$), and
$E_{2qp,{\rm lowest}}$ is the energy of the lowest two-quasiparticle
excitation (with respect to the initial nucleus) corrected for the residual
$pn$ interaction between the valence quasiparticles in the odd-odd system
\cite{jensen}.  The excitation energies of the $1^+$ states with respect to
the final ground state are
\begin{equation}
\label{exc}
E^*_{1^+_m} \approx  E_{\rm QRPA} - E_{2qp,{\rm lowest}}
\quad ,
\end{equation}
where $E_{\rm QRPA}$ is the QRPA phonon energy (see
Sec.~\ref{Subsect:QRPA}).  It follows from Eqs.\ (\ref{eq:binding}) and
(\ref{oneplus}) that the
energy released in the transition from the ground state of the initial nucleus
to a $1^+$ state in the final nucleus is
\begin{eqnarray}
\label{eq:Qvalue}
E_i - E_{1^+_m}
& = & Q_{\beta_-}(Z,N)-E^*_{1^+_m} \nonumber \\
& \approx & \lambda_n - \lambda_p + \Delta M_{n-H} - E_{\rm QRPA}
\quad .
\end{eqnarray}
This expression allows us to avoid calculating ground-state masses of the
final nuclei.

%
%
\subsection{Skyrme-Hartree-Fock-Bogoliubov Method}
\label{Subsect:HFB}

Our first-order description of the even-even ground state and quasiparticle
excitations is based on a self-consistent mean field.  Several schemes are
currently used in mean-field calculations for heavy nuclei, among them the
non-relativistic HF and HFB methods with the Skyrme interaction \cite{SHFrev},
the HFB approach with the Gogny force \cite{Gogrev}, and the relativistic
mean-field model \cite{RMFrev,[Rin96]}.  All these provide a parameterized
energy functional that describes nuclear properties throughout the
chart of nuclides.  But in neutron-rich nuclei near the drip-line, where the
Fermi energy and pairing potential are close in magnitude, pairing cannot be
treated as a small correction to the HF solution.  Instead the physics
underlying the HF and BCS approaches must be incorporated into a single
variational principle, through HFB theory.  Our version of HFB is formulated
in coordinate space and described in detail in Refs.\ \cite{Dob84,Dob96}; here
we discuss only the key ingredients needed for calculating beta decay.

Our
effective interaction comes from the Skyrme energy-density functional ${\cal
E}$.
The functional can be divided into two parts:
\begin{equation}
\label{eq:Efu}
{\cal E}
= {\cal E}_{\rm ph}
  + {\cal E}_{\rm pair}
\quad .
\end{equation}
The particle-hole part ${\cal E}_{\rm ph}$ depends primarily on the density
matrix $\rho_q (\vec{r} \sigma, \vec{r}'\sigma')$, while the pairing
(particle-particle) part ${\cal E}_{\rm pair}$ depends primarily on the
pairing density matrix $\chi_q (\vec{r} \sigma, \vec{r}'\sigma')$ (the index
$q$ denotes protons or neutrons).  The coupling between $\rho$ and $\chi$
comes from the density-dependent terms of the Skyrme and/or pairing
interaction.  Because the nuclei we consider are so neutron-rich, we shall
neglect $pn$ pairing in the HFB ground state (see Sec.~\ref{Sect:inter}); as a
result our HFB wave function will be a product of proton and neutron wave
functions, and matrices of $\rho$ and $\chi$ will be block diagonal in $q$.
Expressions defining the Skyrme and pairing energy functionals have been
presented many times (see e.g.\ Refs.\ \cite{[Vau72],[Eng75],[Bon85],Rei92}),
and so are only briefly discussed here in Secs.\ \ref{Sect:ph-inter} and
\ref{Subsect:likepairing}.

The particle-hole, $h_q (\vec{r} \sigma, \vec{r}' \sigma')$, and
particle-particle, $\Delta_q (\vec{r} \sigma, \vec{r}'\sigma')$, mean fields
are defined as the first derivatives of the energy functional with respect to
the corresponding densities:
\begin{eqnarray}
h_q (\vec{r} \sigma, \vec{r}' \sigma')
& = & \frac{\delta {\cal E}}
           {\delta \rho_q (\vec{r}' \sigma', \vec{r} \sigma)}
      \quad , \\
\Delta_q (\vec{r} \sigma, \vec{r}' \sigma')
& = & \frac{\delta {\cal E}}
           {\delta \chi_q (\vec{r}' \sigma', \vec{r} \sigma)}
      \quad .
\end{eqnarray}
The residual interactions that enter the QRPA equations (discussed in
Sec.~\ref{Subsect:QRPA}) are the corresponding second derivatives of the
energy functional.  Rearrangement terms, which result from the density
dependence of the energy functional
are therefore included in the both the
HFB and the QRPA.

Variation of the energy functional {(\ref{eq:Efu})} with respect to the
densities leads to the coordinate-space HFB equations for protons and neutrons
{\cite{Dob84}},
\begin{eqnarray}
 \int& \! d^3r'& \sum_{\sigma'}
\left(\! \begin{array}{cc}
       h_q (\vec{r} \sigma, \vec{r}' \sigma')  &
       \Delta_q (\vec{r} \sigma, \vec{r}' \sigma') \\
       \Delta_q (\vec{r} \sigma, \vec{r}' \sigma') &
       -h_q (\vec{r} \sigma, \vec{r}' \sigma')
       \end{array}\! \right)\!
\left(\! \begin{array}{c}
       \phi_1 (\alpha, \vec{r}' \sigma') \\
       \phi_2 (\alpha, \vec{r}' \sigma')
       \end{array}\! \right)
    \nonumber \\
  &=&     \left( \begin{array}{cc}
       E_\alpha+\lambda_q & 0           \\
       0           & E_\alpha-\lambda_q
       \end{array} \right)
\left( \begin{array}{c}
       \phi_1 (\alpha, \vec{r} \sigma) \\
       \phi_2 (\alpha, \vec{r} \sigma)
       \end{array} \right) ,  \label{HFB:Eq}
\end{eqnarray}
{where $\alpha$ enumerates the HFB quasiparticle eigenstates.
The  two-component quasiparticle wave functions $\phi_1 (\alpha,
\vec{r}' \sigma')$ and $\phi_2 (\alpha, \vec{r}' \sigma')$
self-consistently define the densities:
}
\begin{eqnarray}
\label{eq:densdef}
\rho_q (\vec{r} \sigma, \vec{r}'\sigma')
& = & \sum_{\alpha \in q} \phi_2 (\alpha , \vec{r}, \sigma) \;
       \phi_2^* (\alpha, \vec{r}', \sigma')
       \quad , \\
\label{eq:chidef}
\chi_q (\vec{r} \sigma, \vec{r}' \sigma')
& = & \sum_{\alpha \in q} \phi_1 (\alpha, \vec{r}, \sigma) \;
      \phi_2^* (\alpha, \vec{r}', \sigma')
      \quad ,
\end{eqnarray}
The Lagrange multipliers $\lambda_q$
 --- the Fermi energies of
the neutrons and protons  --- are fixed by the particle-number conditions
\begin{equation}
N_q
= \int \! d^3 r \sum_{\sigma = \pm}
  \rho_q (\vec{r} \sigma, \vec{r} \sigma)
\quad .
\end{equation}

By working directly in the coordinate space we are able to properly include
unbound states, which, as we have remarked, become important near the neutron
drip line. A particular virtue of our approach is the accurate representation
of the canonical single-particle basis, consisting of eigenstates
$\psi_{q\mu}(\bbox{r}\sigma)$ of the density matrix:
\begin{equation}
                      \int {d}^3{r}' \sum_{\sigma'}
                      \rho_q(\bbox{r}\sigma,\bbox{r}'\sigma')
                      \psi_{q\mu}(\bbox{r}'\sigma')
         = v^2_{q\mu} \psi_{q\mu}(\bbox{r}\sigma) .  \label{eq122a}
\end{equation}
Canonical states form an infinite, discrete, and complete set of localized
wave functions \cite{Dob84,Dob96}; they describe both the bound states and the
positive-energy single-particle continuum.  The canonical basis is well
defined for all particle-bound nuclei, no matter how close they are to the
drip line.  We use it in the next section to formulate the QRPA.

%
%
\subsection{Proton-neutron QRPA}
\label{Subsect:QRPA}
Although one can
derive coordinate-space QRPA equations for the generalized density matrix
\cite{[Mig67],Bor95,Bor96}, here we stick with the older representation in
terms of a discrete single-particle basis.  The canonical HFB basis is ideal
for this purpose because it simplifies the HFB equations, so that HFB + QRPA
can be formulated in complete parallel with BCS + QRPA (described, e.g., in
Ref.\ \cite{eng}).  The $pn$ QRPA equations take the form:
\begin{equation}
\left( \begin{array}{rr}
        A &  B \\
       -B & -A  \end{array} \right)
\left( \begin{array}{c}
    X \\
    Y \end{array} \right)
 = E_{\rm QRPA}
\left( \begin{array}{c}
    X \\
    Y \end{array} \right)
\quad ,
\label{eq:QRPA}
\end{equation}
with matrices $A$ and $B$ defined as
\begin{eqnarray}
A_{pn,p'n'}
& = &   E_{p,p'} \; \delta_{n,n'}
      + E_{n,n'} \; \delta_{p,p'}
      \nonumber \\
&   & + \tilde{V}_{pn,p'n'}  (u_p v_n u_{p'} v_{n'} + v_p u_n v_{p'} u_{n'})
      \nonumber \\
&   & + V_{pn,p'n'} (u_p u_n u_{p'} u_{n'} + v_p v_n v_{p'} v_{n'})
\label{eq:A}
\end{eqnarray}
and
\begin{eqnarray}
B_{pn,p'n'} & = & \tilde{V}_{pn,p'n'} (v_p u_n u_{p'} v_{n'} + u_p v_n
v_{p'}
u_{n'}      \nonumber \\
& & - V_{pn,p'n'} (u_p u_n v_{p'} v_{n'} + v_p v_n u_{p'} u_{n'})  \quad .
\label{eq:B}
\end{eqnarray}
Here $p$, $p'$, and $n$, $n'$ denote proton and neutron quasiparticle
canonical states, $\tilde{V}$ is the $pn$ particle-hole interaction in the 
$1^+$ channel, obtained
from the second derivative of energy functional ${\cal E}$ with respect to the
proton and neutron densities, and $V$ is the corresponding particle-particle
interaction, obtained from the second derivative with respect to the pairing
densities.  The $X$'s and $Y$'s are the amplitudes for exciting
two-quasiparticle and two-quasihole states from the correlated vacuum.  The
occupation amplitudes $v_i$ are the eigenvalues of the density matrix
(\ref{eq122a}).  Since the canonical HFB basis does not diagonalize the HFB
Hamiltonian (\ref{HFB:Eq}), the one-quasiparticle terms in the Eq.\
(\ref{eq:A}) have off-diagonal matrix elements $E_{i,j}$.  The presence of
these terms is the only formal difference between our QRPA equations and those
based on the BCS approximation.  The localized canonical wave functions,
however, are more realistic than the single-quasiparticle states used to mock
up the continuum in the BCS approximation.

To actually solve the matrix QRPA equations (\ref{eq:QRPA}) we have to
truncate the canonical basis.  The diagonal matrix elements of the
one-quasiparticle Hamiltonian provide a convenient measure of excitation, and
we include only those canonical states for which these matrix elements are
less than a cut-off value.  (For the choice of the cut-off energy, see
Sec.~\ref{Sect:inter}.)  Occupation probabilities of canonical states usually
decrease much faster with excitation energy than in the BCS approximation
\cite{Dob96}, allowing us to work with matrices of manageable size.

%
\section{Interactions}
\label{Sect:inter}

Different channels of the effective interaction have different effects on the
beta-strength function.  Roughly speaking, the like-particle and $pn$
particle-hole interaction{s} determine the single-particle spectrum.  The
like-particle pairing interaction smears the Fermi surface and changes the
elementary excitations from particles to quasiparticles.  Without further
many-body correlations, the GT resonance is not collective and too much of the
beta strength is located at low energies.  The $pn$ particle-hole force,
treated in the QRPA, solves the problem by pushing GT strength up and into the
resonance, leaving the low-lying spectrum depleted.  This depletion is so
great, however, that the resulting lifetimes can become too long; the
half-lives of Refs~\cite{mol}, for example, usually exceed measured
half-lives.  The $pn$ particle-particle force cures this problem by pulling
some strength back down.  Interestingly, this part of the force is also
necessary to describe $\beta^+$ decay in proton-rich nuclei, but there it
decreases the low-lying strength.

To be consistent, one should use a single interaction in all channels and in
every step of calculation.  This means, for example, that the same $pn$
particle-hole energy functional that determines single-particle properties in
the HFB calculation must also be used in the QRPA.  The same is true for the
pairing (particle-particle) forces, both in the like-particle and $pn$
channels.  The constraint is not as tight as in the particle-hole channel, 
however, because the proton and neutron Fermi surfaces are so far apart in
neutron-rich nuclei that $pn$ pairing correlations are neglibigle in the HFB.
Furthermore, to the extent that the $T$=1 pairing force
can be approximated by a delta function, it does not affect the
$J^{\pi}$=1$^+$ states obtained in the QRPA.  Even with a finite range force,
the effects are negligible.  The $pn$ component of the $T$=1
pairing/particle-particle interaction can therefore be neglected everywhere.
In addition, one is free to choose the $T$=0 pairing component solely on the 
basis of its effects in the QRPA, since it has no like-particle component and 
does nothing at the mean field level unless $N$$\approx$$Z$.

For these reasons we organize the discussion in this section as follows:
First we describe the particle-hole interaction in detail; it largely
determines both single-particle properties and the collectivity of the GT
resonance.  We then briefly discuss the like-particle pairing force, which
plays a role only in the HFB calculation. Finally, we describe the $pn$
($T$=0) pairing interaction.  This last ingredient appears only in the QRPA
but, as mentioned already, is crucial for reducing calculated lifetimes to
values that are consistent with  experiment.
%

\subsection{The particle-hole interaction}
\label{Sect:ph-inter}

The $pn$ particle-hole interaction is responsible for the main features of the
GT distribution.  For our calculations to make sense, we need to
find an interaction that reproduces the distribution reasonably, at both high
and low energies.  Before doing this, we have to discuss the general form of
the particle-hole interaction in the Skyrme framework.

For even-even nuclei, the particle-hole Skyrme energy functional of
Eq.\ (\ref{eq:Efu}) can be written as
\begin{equation}
{\cal E}_{\rm ph}
  =     {\cal E}_{\rm kin}
      + {\cal E}_{\rm Sk}
      + {\cal E}_{\rm C} \quad ,
\end{equation}
i.e. as the sum of a kinetic-energy functional ${\cal E}_{\rm kin}$,
an effective strong-interaction functional  ${\cal E}_{\rm Sk}$, and a Coulomb
functional ${\cal E}_{\rm C}$.  The functionals are the spatial integrals of
the corresponding local energy densities ${\cal H}$,
\begin{equation}
{\cal E}[\rho, \tau, {\bf J}]
= \int \! {\rm d}^3 r \;
  {\cal H} [\rho({\bf r}), \tau({\bf r}), {\bf J}({\bf r})]
\quad .
\end{equation}
The kinetic-energy and Coulomb energy densities are given by
\begin{eqnarray}
{\cal H}_{\rm kin}
& = & \frac{\hbar^2}{2m} \; \tau \quad ,
      \\
{\cal H}_{\rm C}
& = & \frac{e^2}{2} \int \! {\rm d}^3 r'
      \frac{\rho_{\rm p} (\vec{r}) \rho_{\rm p} (\vec{r}') }
           {|\vec{r} - \vec{r}'|}
      - \frac{3e^2}{4} \left( \frac{3}{\pi} \right)^{1/3} \! \!
        \rho_{\rm p}^{4/3} \quad.
\end{eqnarray}
The Skyrme energy density can be split into pieces ${\cal H}_{\rm Sk}^{\rm
even}$ and ${\cal H}_{\rm Sk}^{\rm odd}$ that are bilinear in time-even and
time-odd densities, respectively \cite{[Eng75],[Dob95]}.  Only
${\cal H}_{\rm Sk}^{\rm even}$ affects the ground-state properties of
even-even nuclei; it can be written as
\begin{eqnarray}
\label{eq:SkyrmeFu}
{\cal H}_{\rm Sk}^{\rm even}
& = &       \frac{b_0}{2}  \rho^2
          + b_1            \rho \tau
          - \frac{b_2}{2}  \rho \Delta \rho
          + \frac{b_3}{3}  \rho^{\alpha +2}
          - b_4            \rho \vec{\nabla} \cdot \vec{J}
      \nonumber \\
&   & - \sum_q \left[
            \frac{b'_0}{2} \rho_q^2
          + b'_1           \rho_q \tau_q
          - \frac{b'_2}{2} \rho_q \Delta \rho_q \right.
      \nonumber \\
&   &
          +\left.  \frac{b'_3}{3} \rho^\alpha \rho_q^2
          + b_4'           \rho_q \vec{\nabla}\cdot \vec{J}_q  \right]
      \nonumber \\
& & - \frac{1}{2}
      \Big[ c_1 \vec{J}^2 - c_1'\sum_q  \vec{J}_q^2 \Big]
      \quad .
\end{eqnarray}
The energy densities depend on the local matter density $\rho_q$, the kinetic
density $\tau_q$, and the spin-orbit current $\vec{J}_q$.  Densities without a
$q$ index are total (isoscalar) densities, e.g.\ $\rho_0 = \rho_p + \rho_n$.
The parameters $b_i$, $b_i'$, $c_1$, and $c_1'$ are the time-even coupling
constants of the Skyrme functional.  The parameters all play different roles;
the terms with $b_3$ and $b_3'$, for example, determine the density-dependent
parts of the interaction, and the $b_4$, $b_4'$ and $c_1$, $c_1'$ terms define
the spin-orbit interaction.  All the parameters are fit to a few key
nuclear-structure data, e.g.\ total binding energies, charge radii, and
surface thicknesses of nuclei in the valley of stability, and they generally
reproduce ground-state properties of nuclei between $^{16}$O and the heaviest
elements \cite{Rei95,SHTB}.

The energy density ${\cal H}_{\rm Sk}^{\rm odd}$, the detailed form of which
can be found in Ref.~\cite{[Dob95]}, involves time-odd
quantities: the momentum density, spin density, and
vector part of kinetic-energy density.  The time-odd terms
contribute to the energy of {\em polarized} states, i.e.\ those with nonzero
angular momentum, including the $1^+$ states populated by beta decay.
As a consequence, the distribution of GT strength depends primarily on this
time-odd part of the energy density.  Although most of the coefficients in
${\cal H}_{\rm Sk}^{\rm odd}$ are in principle independent parameters, they
are fixed by the values of the parameters in ${\cal H}_{\rm Sk}^{\rm even}$
in the usual Skyrme ansatz \cite{[Dob95]}.  The restriction is not in the 
spirit of the local density
approximation \cite{[Neg72],[Neg75]}, but has been made implicitly in almost
all prior work.  The predictions of most existing Skyrme forces for excited
states in odd-odd nuclei are therefore completely determined by fits of ${\cal
H}_{\rm Sk}^{\rm even}$ to properties of even-even ground states.
Not surprisingly, we find the GT spectra predicted by most of these forces to
be well off the mark.

Clearly, the proper approach to beta decay in the Skyrme framework is to
fit coefficients in ${\cal H}_{\rm Sk}^{\rm odd}$
to properties of nuclear states with non-zero angular momentum.
That task is not so easy, however.  Excited states take more
computer time to explore than ground states because they require methods
beyond mean-field theory (such as the QRPA).  Furthermore, ${\cal H}_{\rm
Sk}^{\rm odd}$ contains many parameters, and fixing them all would be a
major undertaking.  For these reasons, we have chosen here to use the
relations between time-even and time-odd densities imposed by the the
traditional Skyrme-force ansatz, in spite of the drawbacks.  (The ansatz,
incidentally, relates the $b_i$, $b_i'$,
$c_1$, and $c_1'$ of Eq.\ (\ref{eq:SkyrmeFu})
to the usual Skyrme-force parameters $t_i$ and $x_i$, as spelled out in
Ref.\ \cite{Rei95}.)  We will explore the effects of independent time-odd
densities in a future publication.
%
%
\mediumtext
\begin{table*}
\caption{Properties of symmetric nuclear matter at saturation density
$\rho_0$ predicted by representative Skyrme interactions.
$E/A$ is the energy per nucleon,
$m^*/m$ the effective mass, $K_\infty$ the incompressibility, and
$a_{\text{sym}}$ the asymmetry coefficient. The remaining quantities are
the Landau-Migdal parameters $f_0, f_0', g_0$, and $g_0'$
\protect\cite{SGII}; $g_0'$ controls the effective interaction in the
spin-isospin channel. The empirical values were taken from
Ref.~\protect\cite{Osterfeld}
}
\label{tab:g0p}
\begin{tabular}{lddddddddd}
Force & $\rho_0$
      & $E/A$
      & $m^*/m$
      & $K_\infty$
      & $a_{\text{sym}}$
      & $f_0$
      & $f_0'$
      & $g_0~~$
      & $g_0'~~$ \\
      & ${[{\rm fm}^{-3}]}$
      & ${[{\rm MeV}]}$
      &
      &
      & ${[{\rm MeV}]}$
      & 
      &
      &
      &   \\ \tableline
SGII & 0.158 & $-$15.58 & 0.786 & 214 & 26.8 & $-$0.233 & 0.728 &    0.622   & 0.934  \\
SkM* & 0.160 & $-$15.75 & 0.789 & 216 & 30.0 & $-$0.229 & 0.926 &    0.325   & 0.937  \\
SkP  & 0.162 & $-$15.93 & 1.000 & 201 & 30.0 & $-$0.102 & 1.417 & $-$0.229   & 0.062  \\
SLy4 & 0.160 & $-$15.97 & 0.694 & 230 & 32.0 & $-$0.276 & 0.813 &    1.385   & 0.901  \\
SLy5 & 0.161 & $-$15.98 & 0.698 & 230 & 32.0 & $-$0.276 & 0.814 &    1.137   & $-$0.152 \\
SLy6 & 0.159 & $-$15.92 & 0.690 & 230 & 32.0 & $-$0.280 & 0.803 &    1.408   & 0.899  \\
SkI4 & 0.160 & $-$15.92 & 0.650 & 248 & 29.5 & $-$0.273 & 0.559 &    1.768   & 0.881  \\
SkO' & 0.160 & $-$15.75 & 0.896 & 222 & 32.0 & $-$0.097 & 1.328 & $-$1.612   & 0.792  \\
Empirical &  &          &       &     &      & 0$\pm$0.3 & $\sim$1.6 & $\sim$0.4 & $\sim$1.80 \\
\end{tabular}
\end{table*}
%
%
\narrowtext

Even with constraints on ${\cal H}_{\rm Sk}^{\rm odd}$, it is 
difficult to include
excited-state properties in the data set used to fit ${\cal H}_{\rm Sk}^{\rm
even}$.  Instead we follow the approach advocated in Ref.~\cite{SGII} and
compare the effective strength of our spin- and isospin-flip interaction in
nuclear matter with phenomenological values or predictions of realistic
calculations.  In nuclear matter the interaction strength in the GT channel is
sensitive to a single combination of the Skyrme parameters, the Landau-Migdal
parameter $g_0'$ \cite{SGII}:
\begin{eqnarray}
\label{Landau}
g_0'
& = & - N_0 \Big[   {\textstyle \frac{1}{3}}
                \Big( b_0 + {\textstyle \frac{1}{2}}  b_0' \Big)
              + {\textstyle \frac{2}{9}}
                \Big( b_3 + {\textstyle \frac{1}{2}}  b_3' \Big)
                \, \rho^\alpha
              + c_1' k_F^2
        \Big] ~,
\end{eqnarray}
where we have used the normalization factor $N_0$ = $2 k_F m^*/(\pi^2
\hbar^2)$, with the Fermi momentum defined as $k_F$ = $(3\pi^2 \,
\rho/2)^{1/3}$.  Table~\ref{tab:g0p} shows predictions of representative
Skyrme interactions -- SGII \cite{SGII}; SkM$^\ast$ \cite{SkM*}; SkP
\cite{Dob84}; SLy4, SLy5 \cite{SLyx}; SkI4 \cite{Rei95}; and
SkO' \cite{Rei99} -- for properties of symmetric nuclear matter at saturation
density.  While all the forces give about the same saturation density
$\rho_0$, energy per nucleon $E/A$, and incompressibility $K_\infty$, small
but noticeable differences appear in the effective mass $m^*$ and asymmetry
coefficient $K_\infty$\footnote{Not all the properties in the Table~1 are
predictions
of the parameter sets; the effective mass of SkP, the
incompressibility of SLy4 and SLy5, and the asymmetry coefficient of
SLy4, SLy5, and SkO' have been used as constraints in the fits.}.
The Landau parameter $f_0$ is a combination of $\rho_0$, $m^\ast$ and
$K_\infty$, and $f_0'$ is related to $\rho_0$ and $m^\ast$;
it is therefore not surprising
that most forces predict similar values for the time-even quantities $f_0$
and $f_0'$, and that those values are close the empirical ones
\cite{Osterfeld}.  But the parameters $g_0$ and $g_0'$, which act in time-odd
channels, reflect spin-dependent components that are not constrained by
standard fits.  As a result their values scatter within a wide range:
\mbox{$-1.6 \lesssim g_0 \lesssim +1.8$} and \mbox{$-0.2 \lesssim g_0'
\lesssim +1$}.  Furthermore, two otherwise very similar forces can give quite
different values for these parameters.  The interactions SLy4 and SLy5, for
example, are identical for one small detail, the treatment
of the $\vec{J}^2$ term during the fit.  But despite the close agreement
for most nuclear-matter parameters, the differences in both $g_0$ and $g_0'$
between the two are significant.

We remarked above that predictions of existing Skyrme forces for GT
distributions usually are not good.  This fact is reflected in
Table~\ref{tab:g0p}
by the values of $g_0'$, all of which are much smaller than the empirical
value \mbox{$g_0' ({\rm expt}) \approx 1.8$} \cite{Osterfeld}.  In fact,
we are not aware of a single Skyrme interaction that gives $g_0'$ close to
the empirical value and at the same time does an acceptable job with
global nuclear properties and single-particle shell structure.
State-of-the-art Skyrme forces tend to yield a $g_0'$ of about 0.9, a
value too small by a factor two.  An important related point:  the value
$g_0'$=0.503 reported for SGII in Ref.~\cite{SGII} does not correspond to
the consistent application of any Skyrme force.  The authors left out the
last term (in brackets) in Eq.\ (\ref{eq:SkyrmeFu}) when fitting the
parameters of their force, but included its effects in their calculation
of $g_0'$ (and in their RPA calculations).  To be consistent with their
mean-field interaction, they should have omitted the term proportional to
$k_F^2$ in Eq.\ (\ref{Landau}).  Doing so gives $g_0'$=$0.93$, the value
in Table~\ref{tab:g0p}.  We will elaborate on this remark in a future
paper. For now, what's important is that even calculated properly the
value of
$g_0'$ associated with this interaction, which was designed
explicitly for GT resonances, is far too small.

\begin{figure}[t]
\centerline{\epsfig{file=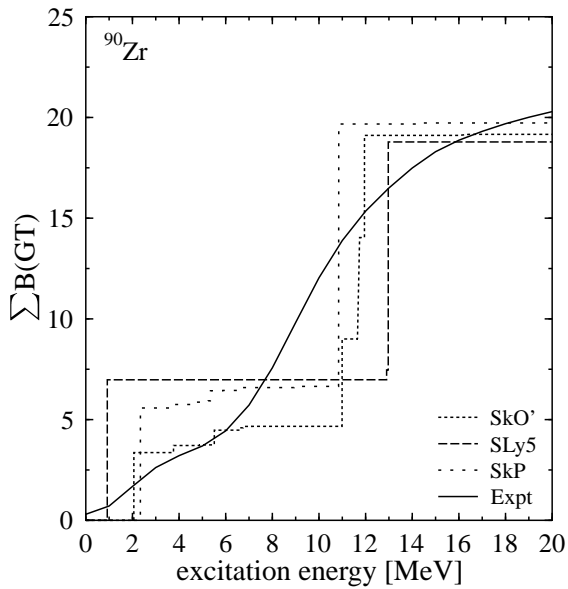}}
\caption{\label{GT90Zr}
Summed GT strength up to 20 MeV as a function of excitation energy for
the closed-shell nucleus $^{90}$Zr,
calculated with the SkO', SLy5, and SkP Skyrme forces. We also plot the
measured strength reported in Ref.\ \protect\cite{Japanese} The calculated
strength, as is customary, is multiplied by $(1/1.26)^2$; the quenching
corresponds to setting $g_A$ to 1.0 in our calculations of beta decay.
}
\end{figure}

\begin{figure}[t]
\centerline{\epsfig{file=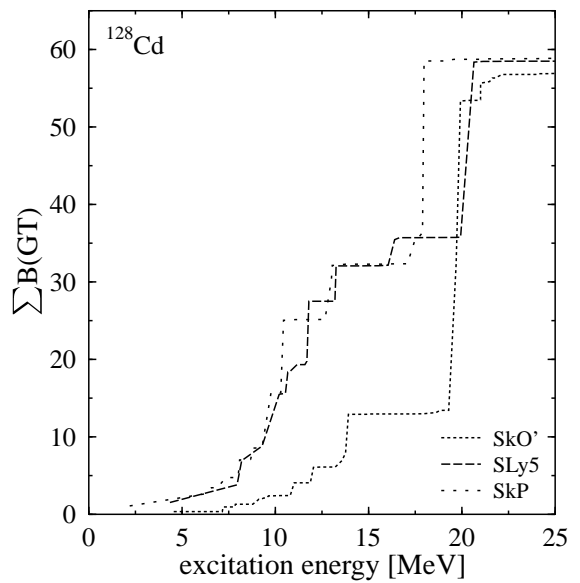}}
\caption{\label{GT128Cd}
Same as Fig.~\protect\ref{GT90Zr}, but for
the  open-shell nucleus $^{128}$Cd. No experimental data exist here.
}
\end{figure}

Of course $g_0'$ in symmetric nuclear matter at saturation density does
not directly measure the strength of spin- and isospin-flip interactions
in a finite nucleus.  All effects due to the nuclear surface, finite
nuclear volume, and excess neutrons disappear in nuclear matter.
Moreover, the GT distribution depends strongly on the single-particle
spectrum as well as the residual interaction.  Nonetheless, the parameter
$g_0'$ summarizes the gross features of the distribution. This is
illustrated in Figs.~\ref{GT90Zr} and \ref{GT128Cd}, which display the
summed GT strength calculated for $^{90}$Zr and $^{128}$Cd.  Here we use
three Skyrme forces with different values of Landau-Migdal parameter:
SkO' ($g_0'$=0.792), SkP ($g_0'$=0.062), and SLy5 
($g_0'$=--0.152).  In
general both the energy of the GT resonance and the strength in it
increase with $g_0'$; they are smallest for SLy5 and largest for SkO'.
Forces with small values of $g_0'$ fail to concentrate enough strength in
the resonance, leaving too much at low energy.

Although none of the interactions have large enough $g_0'$ when used
self-consistently, some are better than others.  Having opted for now not to
increase $g_0'$ by using energy functionals in which 
${\cal H}^{\rm even}_{\rm Sk}$ and ${\cal H}^{\rm odd}_{\rm Sk}$
decoupled, we decided to use SkO' \cite{Rei99}, which has one of the 
larger values in Table~\ref{tab:g0p}.
The SkO' energy functional is more general than the
original Skyrme functionals in the spin orbit channel;
the extra generality manifests itself as a difference in the
values of $b_4$ and $b_4'$ \cite{Rei95}. Without this extended interaction it
seems to be impossible to obtain a consistent description of spin-orbit
splittings and other global observables unless the last term in brackets in
Eq.\ (\ref{eq:SkyrmeFu}) is
neglected \cite{Rei99a}.  The extension remedies
the problem quite nicely, and in spite of the low value of $g_0'$, SkO'
reproduces GT spectra \cite{rap} measured in charge-exchange reactions fairly
well.  The predicted resonances, though usually a little too low in energy,
contain about the correct fraction of the strength\footnote{This observation
is important because most published Skyrme-RPA calculations do not address the
problem of the GT strength {\em distribution}.  For instance, Ref.~\cite{Colo}
proposes that the difference between the centroid energy of the GT resonance
and the energy of the isobaric analog state be included as a constraint on
effective interactions.  Getting the right energy difference seems to be too
weak a criterion; the force SGII, which passes the test in Ref.~\cite{Colo},
has a value of $g'_0$=0.93 that is well below the empirical one.}.
%
%
%
%
%
%
%
%
%
%
\subsection{Like-particle pairing interaction}
\label{Subsect:likepairing}
The $T$=1 pairing interaction between like particles affects only the HFB part
of our calculation.  To incorporate it we use a simple pairing energy
functional that corresponds to a delta force,
\begin{equation}
{\cal E}_{\rm pair}
= \frac{1}{4} \int \! d^3 r \; \sum_{q = p,n} V_{1,q} \chi_q^2 (\vec{r})
\quad ,
\end{equation}
where \mbox{$\chi_q(\vec{r}) = \sum_\sigma \chi_q(\vec{r}, \sigma;
\vec{r}, \sigma)$}  is the local  pairing tensor.
This interaction has vanishing matrix elements in the $1^+$ channel and
therefore contributes nothing to the residual interaction in the $pn$
QRPA.  We adjust the strengths $V_{1,q}$ in the HFB calculation to reproduce
experimental pairing gaps as explained in Ref.\ \cite{Dob95}, though unlike
the authors of that paper we use different values for protons and
neutrons, and allow them to depend slightly on mass.
This refinement appears to
be necessary for a precise description of the beta decay rates.
For light nuclei with $N \approx 50$ we adopt the values
\begin{displaymath}
V_{1,p} = -188.1 \; {\rm MeV} \; {\rm fm^3}
\quad , \quad
V_{1,n} = -213.8 \; {\rm MeV} \; {\rm fm^3}
\quad ,
\end{displaymath}
while for heavier nuclei with $N \approx 82$ we use
\begin{displaymath}
V_{1,p} = -194.6 \; {\rm MeV} \; {\rm fm^3}
\quad , \quad
V_{1,n} = -186.7 \; {\rm MeV} \; {\rm fm^3}
\quad .
\end{displaymath}
%
%
%
%
\begin{figure}[t]
\centerline{\epsfig{file=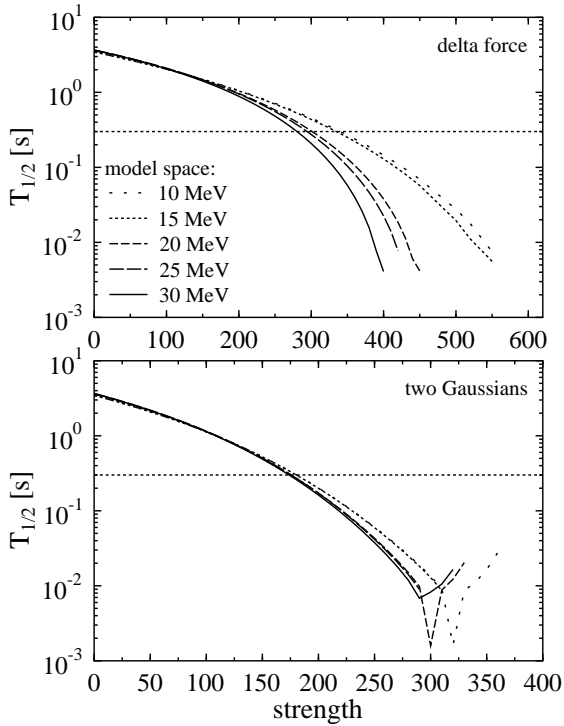,width=7.5cm}}
\caption{\label{fig:cd128_modelspace}
Dependence of the half-life of $^{128}$Cd on the size of the
canonical basis used in the QRPA, plotted versus
the strength of the $T$=0 pairing interaction. The lower and upper
panels show results obtained with the finite-range force of
Eq.~(\protect\ref{eq2}) and a delta force.
The configuration  space in QRPA is defined by the cut-off energy, $e_{\rm
cutoff}$ as described in the text.
The horizontal line shows the experimental half-life of $^{128}$Cd.
}
\end{figure}
%
%
\subsection{Residual $pn$ particle-particle interaction}
\label{Subsect:ppres}
Although the GT resonance is built almost entirely of particle-hole
excitations, the low-lying strength responsible for beta decay involves $pn$
particle-particle correlations and is sensitive to the $T$=0 pairing
interaction.  The reason these correlations are important at low energies is
that the proton orbitals near the Fermi surface are neither completely empty
nor completely occupied.  They therefore can accept the additional particle
created from occupied neutron orbitals by beta decay at the same time as they
interact with those neutron orbitals through the $T$=0 pairing force.  A level
that is completely full, by contrast, can interact with the occupied neutron
orbitals but will not participate in beta decay, while one that is completely
empty can accept additional protons from beta decay but will experience no
particle-particle interaction with the occupied neutron levels.

Because $T$=0 pairing has no effect in our HFB calculations, we can treat its
strength as a free parameter in the QRPA.  Our procedure is to fit that
strength to known beta-decay lifetimes in regions of the nuclear chart near
those that interest us.  To keep matters simple, we restrict ourselves to a
density-independent force acting only in the $S$=1 channel; QRPA calculations
of double-beta decay \cite{eng}
and single-beta decay \cite{Muto,Homma}
have shown this component of the
interaction to have the largest effect on low-lying strength.  Unfortunately,
although a simple delta force successfully describes like-particle pairing, it
turns out to be inadequate here; the calculated half-lives diverge steadily as
canonical single-quasiparticle states are added to the basis used in the QRPA.
Actually, any purely attractive interaction suffers from the same problem.
The situation improves considerably, however, when we use (as a crude mock up
of a microscopic $G$-matrix) a short-range repulsive Gaussian combined with a
weaker longer-range attractive Gaussian:
\begin{equation}
\label{eq2}
V_{12}
= - V_0 \sum_{j=1}^2 g_j \; {\rm e}^{-r_{12}^2/\mu_j^2} \;
    \hat\Pi_{S=1,T=0}
\quad ,
\end{equation}
%
%
%
\begin{figure}[t!]
\centerline{\epsfig{file=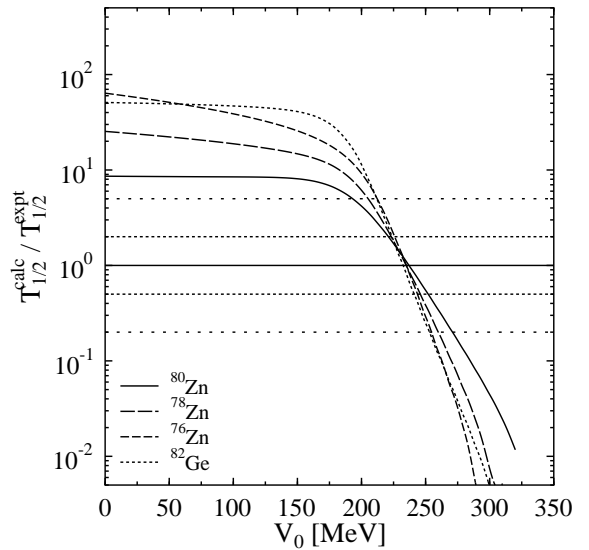,width=7.5cm}}
\caption{\label{fig:zn_v0}
Ratios of calculated-to-measured half-lives for four nuclei near
\mbox{$N=50$} as a function of the $T$=0 paring strength.  The
solid horizontal line corresponds to equal values for
measured and calculated half-lives, the short-dashed horizontal line to a
factor of 2 difference, and the dotted line to a factor of 5 difference.
}
\end{figure}
%
%
%
%
\begin{figure}[h!]
\centerline{\epsfig{file=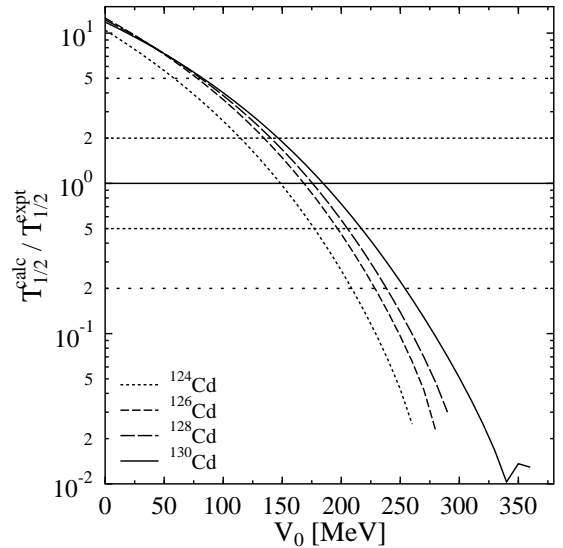,width=7.5cm}}
\caption{\label{fig:cd_v0}
Same as figure~\protect\ref{fig:cd_v0}, but for nuclei near
\mbox{$N=82$}.
}
\end{figure}
%
%
%
%
\widetext
\begin{figure*}[t!]
\centerline{\epsfig{file=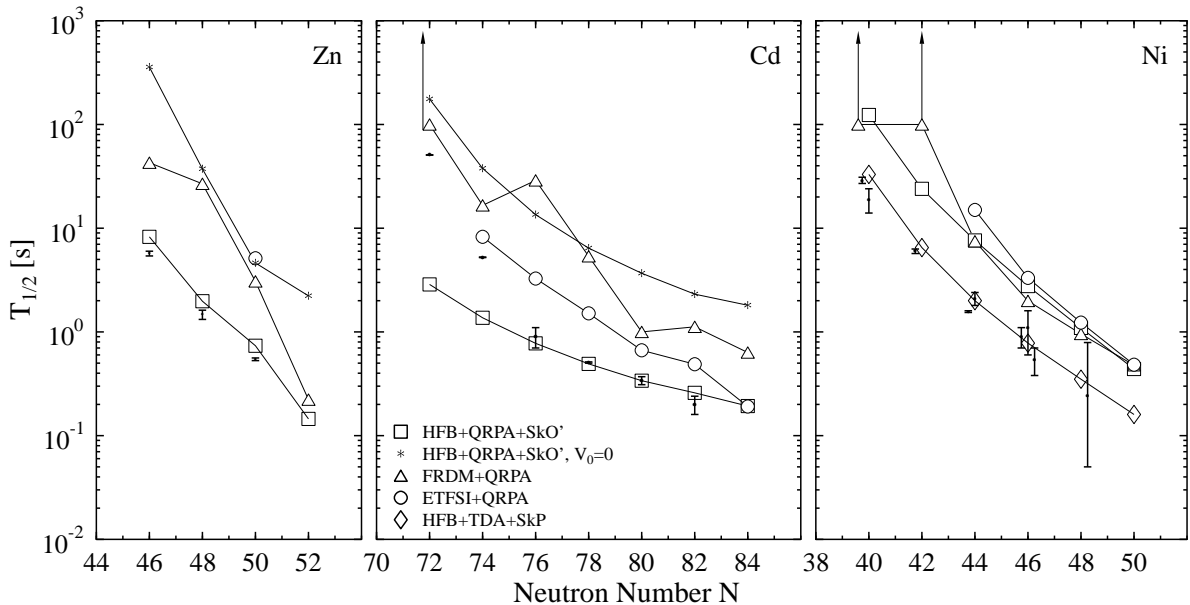}}
\caption{\label{fig:hl:fit}
Comparison of calculated half-lives for Zn, Cd, and Ni isotopes
with (HFB+QRPA+SKO') and without (HFB+QRPA+SkO', \mbox{$V_0 = 0$}) the
residual particle-particle interaction, and results from Refs.\
\protect\cite{mol} (FRDM+QRPA), \protect\cite{Bor97} (ETFSI+QRPA),
and \protect\cite{zyl} (HFB+TDA+SkP, only for the Ni isotopes), with 
experimental values taken from Ref.\ \protect\cite{TOI} where possible. For 
the nickel isotopes recent results from \protect\cite{ameil,Fra98} are shown
as well.  When predicted half-lives are larger than 100 s, the FDRM
collaboration reports only this lower bound, which is marked here with arrows 
pointing up. The ETSFI collaboration reports half-lives only when the 
predicted deformation $\beta_2$ is less that 0.1.
}
\end{figure*}%
\narrowtext%
%
%
where $\hat\Pi_{S=1,T=0}$ projects onto states with $S=1$ and $T=0$.  We take
the ranges $\mu_1$=1.2\,fm and $\mu_2$=0.7\,fm of the two Gaussians from the
Gogny interaction \cite{Gogny}, and choose the relative strengths $g_1$=1 and
$g_2$=--2 so that the force is repulsive at small distances.  The only
remaining free parameter is $V_0$, the overall strength.
Figure~\ref{fig:cd128_modelspace} depicts the rate of convergence of the
predicted half-life of $^{128}$Cd as a function of $V_0$ for the interaction
(\ref{eq2}) and for a $\delta$ interaction.  The latter gives results that do
not converge when we increase $e_{\rm cutoff}$, the upper limit on the
(diagonal) energy of canonical states we include in the QRPA.  The
finite-range force behaves much better.  The calculations reported below were
carried out with the two-Gaussian\footnote{Other choices are possible.  For
instance, we have checked that the convergence can also be achieved with a
simplified particle-particle interaction of Skyrme type with $t_0 = 6.1697 \,
V_0 \, {\rm MeV} \; {\rm fm}^3$, $t_1 = -2.5849 \, V_0 \, {\rm MeV} \; {\rm
fm}^5$, $x_0=x_1=0.25$, and all others equal to zero.}  force in Eq.\
(\ref{eq2}) and with $e_{\rm cutoff}$=25\, MeV for states above the Fermi
surfaces (we include all states below the Fermi surfaces).

To fit the strength $V_0$, we use recently measured half-lives of neutron-rich
nuclei in regions where the r-process path comes closest to the valley of
stability (e.g., around $^{78}$Ni) \cite{engelmann,ameil,Fra98}). 
Fig.~\ref{fig:zn_v0} displays the ratio of calculated-to-experimental
lifetimes for three zinc isotopes and $^{82}$Ge, all near $N$=50.  The four
lines intersect near a ratio of 1, showing that the experimental lifetimes
can all be reproduced with a single value of the $T$=0 pairing strength,
$V_0$=230 MeV.  We use this interaction strength to predict the lifetimes of
nuclei (in this region) that are still further from stability.

Fig.~\ref{fig:cd_v0} shows the corresponding results for $^{126,128,130}$Cd,
all near $N$=82.  The fit, while not as good here, is still reasonable and
gives a best value for $V_0$ of about 170 MeV.  At that value we slightly
underestimate the lifetime of $^{124}$Cd (by a factor of 0.6).  For $^{122}$Cd
the discrepancy is greater --- we underestimate its lifetime by a factor of 5
--- but that nucleus has such a small $Q_{\beta^-}$--value that a small error
in the strength distribution can have a large effect on the rate.  We
therefore adopt the value $V_0$=170 for the entire region.  The reason the
value of $V_0$ is so different here might be connected with the quality of the
single-particle spectra and $Q$-values predicted by Skyrme forces, and with
the simplicity of our $T$=0 pairing interaction.  We may be compensating for
such deficiencies by changing $V_0$.  The fact that we have to change the
$T$=0 pairing strength by such a large amount when going from $N$=50 to $N$=82
demonstrates the sensitivity of calculated beta decay rates to other parts of
the effective interaction.

In the $N$=126 region there are not enough experimental data to support a fit
of the $T$=0 pairing strength, so we use the same value ($V_0$=170\,MeV) as in
the $N$=82 region.  This means that our predictions around $N$=126 are less
reliable than in other regions (see, however, discussion in
Sec.~\ref{Sect:results}).
%
%
\section{Results}
\label{Sect:results}
%
%
%
%
\begin{figure}[t!]
\centerline{\epsfig{file=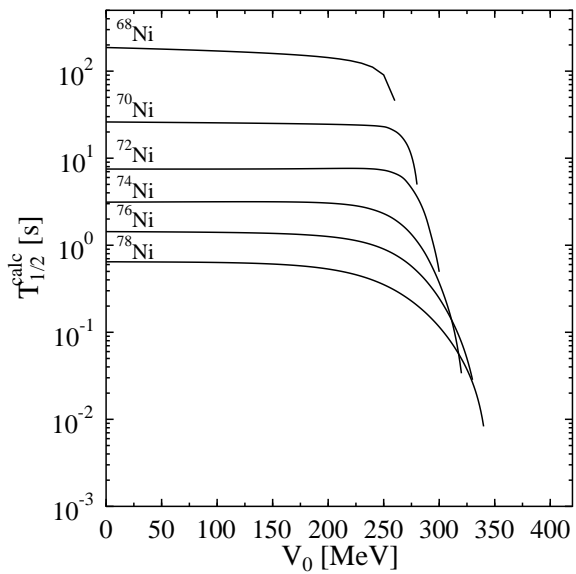}}
\caption{\label{fig:ni_v0}
Calculated half-lives of the neutron-rich Ni isotopes as a function of the 
strength of the $T$=0 pairing interaction.
}
\end{figure}
%
%
\subsection{Measured half-lives}
\label{Sect:results:A}
Figure~\ref{fig:hl:fit} displays the results of our calculations for nuclei
near the neutron closed shells, alongside the QRPA results without $T$=0
pairing ($V_0$=0), the results of Refs.\ \cite{Bor97,zyl,FRDMdata}, and
measured half-lives \cite{TOI}.  (The predictions of Ref.~\cite{Bor97} appear
only when the nucleus is thought to have small deformation.  Our numbers in
these nuclei are more suspect than elsewhere, since we ignore deformation in
our calculation.)  Because we separately adjust the $T$=0 pairing strength in
two regions, it is not surprising that our results are usually closer to
experiment than those of global calculations in which parameter values are
kept fixed.  But the errors in the global calculations are systematic; they
almost always overestimate the lifetimes.  In the case of Ref.\
\cite{mol,FRDMdata}, at least, we attribute the problem to the neglect of the
$T$=0 pairing.  As the figure shows, the results of Ref.\ \cite{mol,FRDMdata}
are much closer to ours in most of the nuclei when we turn that force off.

In the Ni isotopes our lifetimes are also too large by factors of 3-5.  Part
of the reason is the weak sensitivity of these lifetimes to $V_0$, which in
turn is due to the $Z$=28 and $N$=40 shell closures.  The proton orbitals
important for beta-decay are completely empty and therefore couple through
$T$=0 pairing only to neutron orbitals that are at least partly empty.
Because of the closed shell at $N$=40, the lowest such neutron orbit is
$g_{9/2}$.  But the $g_{9/2}$ neutrons can only decay into the proton $g$
orbitals, which are far above the proton Fermi surface.  Thus no low-energy
strength can be moved by the particle-particle interaction and the curves in
Fig.\ \ref{fig:ni_v0}, which shows the calculated half-lives of the nickel
isotopes versus $V_0$, are all flat.  They do not turn down until after the
value of $V_0$ (230 MeV) that is suitable for the other nuclei in the region,
and close to the point at which the QRPA collapses.  Such behavior clearly
means that SkO' is not optimal in this region; the SkP interaction used in the
calculations of Ref.\ \cite{zyl} apparently better predicts single-particle
properties and Q-values.  Those calculations use no particle-particle force
but, as discussed above, none is required in the Ni isotopes.

Our HFB + QRPA theory violates particle number conservation, with the result
that pairing correlations artificially break down at closed shells.  The
predictions for nickel would probably be better in a number-conserving version
of our approach (for the general formalism, see Ref.~\cite{RingQRPA}).
%
%
\begin{figure}[t!]
\centerline{\epsfig{file=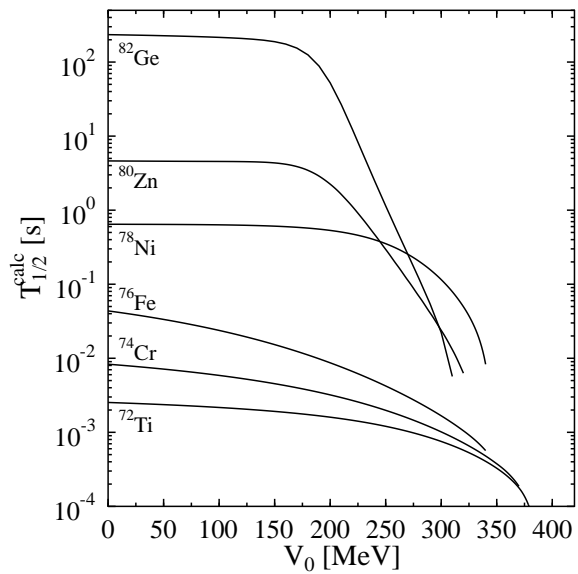}}
\caption{\label{fig:n50_v0}
The same as figure~\protect\ref{fig:ni_v0}, but for \mbox{$N=50$} isotones.
}
\end{figure}
%
%
%
%
\subsection{Closed-neutron-shell r-process waiting points}
The effects of $T$=0 pairing vary just as much in closed-neutron-shell nuclei
along the r-process path as they do in the measured nuclei just discussed.  In
doubly magic nuclei like ${}^{78}$Ni, ${}^{132}$Sn, and ${}^{122}$Zr, the
$T$=0 pairing force is ineffective.  On the other hand, when one takes away
two protons from these nuclei, creating two holes in a high-spin orbit, the
force has a large effect; the corresponding neutron orbit and its spin-orbit
partner are not too far below the neutron Fermi surface and contain many
neutrons, which both interact with the protons at their Fermi surface and
decay to fill the two proton holes.  The effect of adding two protons to a
closed shell is a bit smaller.  The several orbits above the closed proton
shell have lower spin or are far from the Fermi surface, and their
contributions tend to cancel.  These points are illustrated in
Fig.~\ref{fig:n50_v0}, which shows the dependence of the calculated half-lives
of several \mbox{$N=50$} isotones on $V_0$.  The half-life of the doubly magic
nucleus ${}^{78}$Ni, of course, varies almost not at all, and we probably
overpredict its lifetime slightly just as in the other Ni isotopes.
%
%
\widetext
\begin{figure*}[t!]
\centerline{\epsfig{file=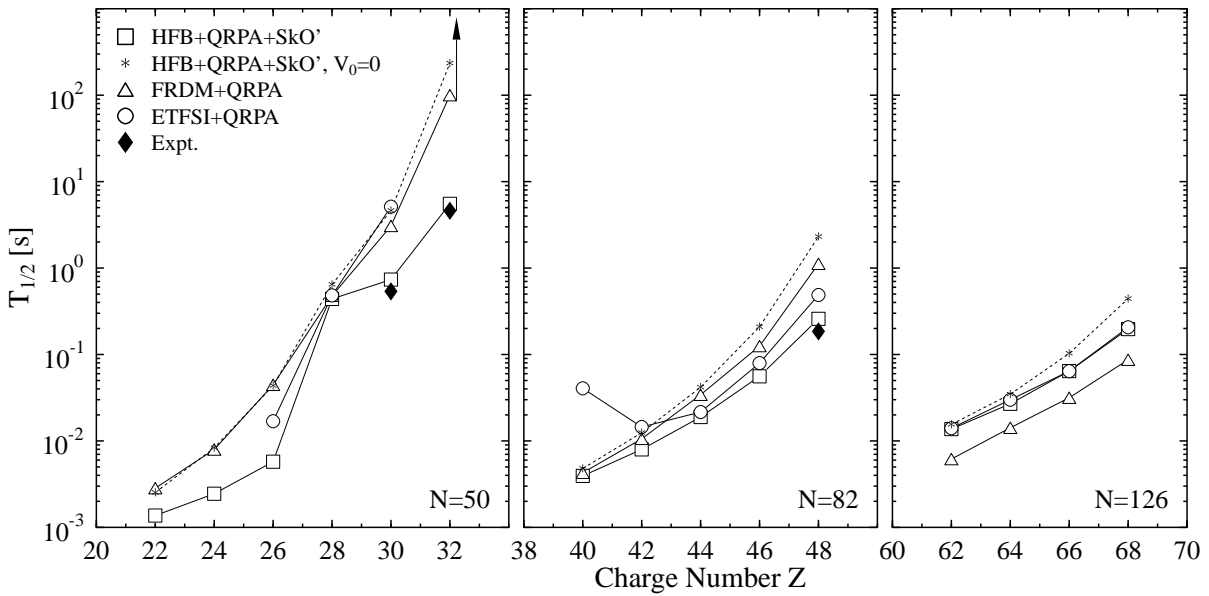}}
\caption{\label{fig:hl:rprocess}
Predictions for the half-lives of closed neutron-shell nuclei along the
r-process path.  Our results appear with (HFB+QRPA+SKO') and without 
(HFB+QRPA+SkO', \mbox{$V_0 = 0$})the $pn$ particle-particle interaction.  Also 
plotted are the results of Ref.\ \protect\cite{mol} (FRDM+QRPA), Ref.\ 
\protect\cite{Bor97} (ETFSI+QRPA), and experimental data where available.
}
\end{figure*}
\narrowtext
%
%

Figure \ref{fig:hl:rprocess} shows our predictions, together with those of
other authors, for the half-lives of all the crucial even-even
closed-neutron-shell nuclei along the r-process path.  Our results agree
fairly well with those of Ref.\ \cite{mol} for the very proton-poor nuclei
(with \mbox{$N=50$} and \mbox{$N=82$}) but less well for larger $Z$.  The
trend is due to the closed proton shells at $Z=20$, 28, 40, and 50, where the
particle-particle force has little effect.  Between these magic numbers, and
particularly just below them (e.g., in ${}^{76}{\rm Fe}$), the differences can
be large.  To demonstrate again that they are due to $T$=0 pairing, we plot
results once more with that component of the force switched off ($V_0$=0), a
step that brings our results into agreement with those of Ref.\ \cite{mol} in
nearly all nuclei with $N$=50 or 82.

As discussed in Sect.\ \ref{Subsect:ppres}, there are no experimental data
with which to fix $V_0$ near $N$=126.  The lack of closed shells in this
region suggests that half-lives will depend strongly on $V_0$.  Our results
with and without $T$=0 pairing, however, show that this is not the case.  Even
if we used much a smaller value of $V_0$, by extrapolating the drop in that
parameter between $N$ = 50 and 82, the lifetimes would not change appreciably.
In these heavy systems our results agree well with those of Ref.\
\cite{Bor97}.
%
%
\subsection{Consequences for nucleosynthesis}
The closed-neutron-shell nuclei are instrumental in setting abundances
produced in the r-process; new predictions for their half-lives will have an
effect on the results of r-process simulations.  For $N$=$50$ and 82 our
half-lives are usually shorter than the commonly employed half-lives of Ref.\
\cite{mol}, and longer for $N$=$126$ Replacing those lifetimes with ours
should therefore produce smaller $A \approx 80$ and 130 abundance peaks, and a
larger $A \approx 195$ peak.

Without extending our calculations to other nuclei in the r-process network,
however, we cannot draw quantitative conclusions from a simulation.
Accordingly, we carry out only one simple r-process simulation here, comparing
final abundance distributions obtained from the beta decay rates of Ref.\
\cite{mol} with those obtained from our
calculations, leaving all other ingredients
unchanged (We also change rates at $N$=84 and 86, by amounts equal to the
change in corresponding nuclei with $N$=82.)  By specifying an appropriate
temperature and density dependence on time, we mock up conditions in the
``neutrino-driven wind" from Type II supernovae, the current best guess for
the r-process site.

\begin{figure}[t!]
\centerline{\epsfig{file=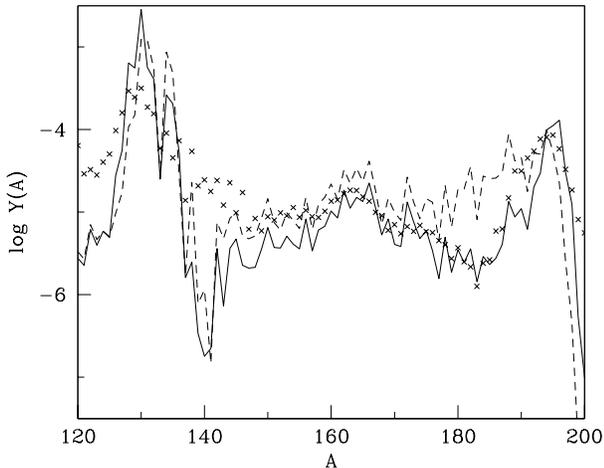,width=8cm}}
\caption{\label{fig:abundances}
Predicted abundances in a simulation of the r-process.  The solid line
corresponds to the rates of Ref.\ \protect\cite{mol}, and the dotted line to
the rates obtained here around $N$=82 and 126. All other
nuclear and astrophysical parameters are the same for the two lines. The
crosses are observed solar-system abundances.
}
\end{figure}

The results appear in Fig.\ \ref{fig:abundances}.  As expected, the $A \approx
130$ peak shrinks noticeably.  The $A \approx 195$ peak broadens with the new
half-lives because abundances around $N$=126 are built up not just at the
longest lived (most stable) nucleus produced, but at more neutron-rich $N$=126
nuclei as well.  As a result, more nuclei are populated and the peak widens.
By contrast, the $N$=50 peak (not shown in Fig.~\ref{fig:abundances}) does not
change much; its shape depends largely on the half life of $^{78}$Ni, which in
our calculations is almost the same as in those of Ref.~\cite{mol}.  We have
already pointed out, however, that both lifetimes are probably too long.  It
is therefore reasonable to expect larger changes at low $A$ than our
simulation indicates.

Besides altering the distribution of abundances, smaller half-lives can
shorten the time required for the r-process to synthesize all the elements.
The process proceeds only as fast as material can move through ``bottlenecks",
the especially long-lived isotopes at the three closed neutron shells.  In
many simulations the sum of the lifetimes of the bottleneck nuclei
exceeds the expected duration of r-process conditions in neutrino-driven winds
\cite{qian}.  To see what our lifetimes do, we run a series of r-process
simulations at different temperatures, with both the half lives of Ref.\
\cite{mol} and with those presented here.  At low temperatures, for which
neutron photodissociation rates are slower, the average nucleus is extremely
neutron rich and so the change in beta half-lives does not have a large
effect; the time required for the r-process drops by $\approx 15\%$.  At
higher temperatures, nuclei that are slightly less neutron-rich are produced,
and our half-lives have a more significant impact, resulting in an r-process
time about $50\%$ shorter.  These high-temperature simulations have difficulty
reproducing the observed abundance distribution, so we cannot at present take
them very seriously.  It might be possible, however, to alter other conditions
so that the correct distribution is restored.  In any case, quantitative
insight will have to await more comprehensive calculations of beta decay.
%
%
%
%
%
%
\section{Summary and conclusions}
\label{Sect:summary}

This work contains the first consistent application of the coordinate-space
HFB+QRPA formalism to beta decay in heavy neutron-rich
nuclei.  Our approach fully accounts for the coupling between the
nuclear mean field, pairing, and the particle continuum.  In addition our
results are based on {\em one} Hamiltonian; we use the same interaction,
SkO', in the HFB and the QRPA calculations.

Beta-decay half-lives depend on Q-values, shell structure, and the residual
interaction.  Much of the effect of the last ingredient is summarized by the
Landau-Migdal parameter $g_0'$.  Apparently all Skyrme interactions commonly
used in nuclear structure studies, including SkO', have values of $g_0'$ that
fall well below the experimental estimate ($g_0'$$\sim$1.8).  As a result, the
centroid of the GT strength distribution is usually too low.  That this defect
is not fatal is due to pairing.  Like-nucleon pairing produces diffuse Fermi
surfaces, which in turn allow $T$=0 pairing to pull strength down in energy.
We can therefore compensate for a small $g_0'$ by weakening the strength of
the $T$=0 pairing interaction slightly .  But we definitely cannot omit it
altogether; as Fig.~\ref{fig:hl:fit} shows, calculations that do almost always 
overestimate lifetimes.

As for the r-process itself, shorter beta-decay rates will have an impact, but
just how much is not yet clear.  Abundances certainly shift somewhat, and the
time it takes to complete the r-process will shrink, perhaps substantially.
But in order to explore these issues fully, one needs to know half-lives for
all the waiting point nuclei, not just those at the closed shells.  Other
quantities that affect abundance distributions, particularly neutron
separation energies, also have to be better understood \cite{Chen95,Pfe97}.

One virtue of our self-consistent framework is that many extensions and
improvements can be made systematically.  First, a new interaction, based on
the concept of the energy density functional, can be developed.  This force
should be able to reproduce global nuclear observables, and at the same time
yield a value of $g_0'$ that is close to 1.8.  Finding such a force will
require abandoning some of the conventional relations between the ${\cal
H}^{\rm even}_{\rm Sk}$ and ${\cal H}^{\rm odd}_{\rm Sk}$ components of the
energy density \cite{[Dob95]}.  Then the parameters of ${\cal H}^{\rm
even}_{\rm Sk}$ can be adjusted to global nuclear properties while those of
${\cal H}^{\rm odd}_{\rm Sk}$ can be fit to spin-dependent properties ($g_0$,
$g_0'$, moments of inertia, etc.).

Following the development of a better interaction, one can both improve the
calculations presented here and extend them to open-shell nuclei that are not
spherical.  The formalism might also to be developed so that particle-number
and isospin conservation are at least partly restored.  Finally, all the
nuclear ingredients in r-process network simulations, including
neutron-separation energies, should be based on the same effective
Hamiltonian.  Only then might our understanding of neutron rich nuclei be
sufficient to predict details of the r-process abundance distribution.  This
paper is a first step towards that goal.

%
%
\section*{Acknowledgments}
The authors thank P.--G. Reinhard for useful suggestions and for the
parameters of the SkO' interaction, H. Sakai for supplying us with 
experimental data on the GT distribution in ${}^{90}$Zr, and I. Borzov, 
S. Fayans, and P. Vogel for discussions. This work was supported in part 
by the U.S.\ Department of Energy under Contract No{s}.\ 
DE--FG02--97ER41019 {(}University of North Carolina{)} and 
DE--FG02--96ER40963 {(}University of Tennessee{)}, and
by the Polish Committee for Scientific Research (KBN) 
under Contract No.~2~P03B~040~14.
%
%

\end{document}